\documentclass[pdflatex,sn-mathphys-num]{sn-jnl}

\usepackage{graphicx}%
\usepackage{multirow}%
\usepackage{amsmath,amssymb,amsfonts}%
\usepackage{amsthm}%
\usepackage{mathrsfs}%
\usepackage[title]{appendix}%
\usepackage{xcolor}%
\usepackage{textcomp}%
\usepackage{manyfoot}%
\usepackage{booktabs}%
\usepackage{algorithm}%
\usepackage{algorithmicx}%
\usepackage{algpseudocode}%
\usepackage{listings}%
\usepackage{placeins}
\usepackage{mathrsfs}

\theoremstyle{thmstyleone}%
\theoremstyle{thmstyletwo}%

\theoremstyle{thmstylethree}%

\begin{document}

\title[Article Title]{Reinforcement Learning-Enabled Dynamic Code Assignment for Ultra-Dense IoT Networks: A NOMA-Based Approach to Massive Device Connectivity}








\author*[1]{\fnm{Sumita} \sur{Majhi}}\email{sumit176101013@iitg.ac.in}
\equalcont{These authors contributed equally to this work.}

\author[1]{\fnm{kishan} \sur{Thakkar}}\email{t.kishan@alumni.iitg.ac.in}
\equalcont{These authors contributed equally to this work.}

\author[1]{\fnm{Pinaki} \sur{Mitra}}\email{pinaki@iitg.ac.in}

\affil[1]{\orgdiv{Department of Computer Science and Engineering}, \orgname{Indian Institute of Technology Guwahati}, \orgaddress{\city{Guwahati}, \postcode{781039}, \state{Assam}, \country{India}}}




\abstract{Ultra-dense IoT networks require an effective non-orthogonal multiple access (NOMA) scheme, yet they experience intense interference because of fixed code assignment. We suggest a reinforcement learning (RL) model of dynamic Gold code assignment in IoT-NOMA networks. Our Markov Decision Process which is IoT aware is a joint optimization of throughput, energy efficiency, and fairness. Two RL algorithms are created, including Natural Policy Gradient (NPG) to learn stable discrete actions and Deep Deterministic Policy Gradient (DDPG) with continuous code embedding. Under smart city conditions, NPG can attain throughput of 11.6\% and energy efficiency of 15.8 likewise superior to its performance with a static allocation. Nonetheless, the performance is worse in organized industrial settings, and the reliability is minimal (0--2\%), which points to the fact that dynamic code assignment is not a sufficient measure of ultra-reliable IoT and needs to be supplemented by power control or retransmission schemes. The work offers a basis to the RL-based resource allocation in massive IoT network.}


\maketitle

\section{Introduction}
\label{sec:introduction}

Internet of Things (IoT) is fast becoming an ubiquitous cyber-physical fabric, and is predicted to reach over 75 billion connected devices by 2025 \cite{mekki2023}. New uses like smart cities \cite{mehmood2017internet} and industrial automation are compelling deployments into ultra-dense networks with more than one million devices per square kilometer~\cite{mehmood2017internet}. This congestion, together with the non-uniform traffic, energy, and various Quality-of-Service (QoS) demands, brings forward drawbacks to conventional Orthogonal Multiple Access (OMA) schemes\cite{guo2021enabling}.

NOMA has become a major facilitator of massive IoT in beyond-5G and 6G systems \cite{dai2018} where devices may share time-frequency resources by multiplexing power or the code-domain. Code-domain NOMA is especially appealing to IoT because it has no grants and can be used to access asynchronous channels \cite{ding2017survey}. In very dense environments, however, it has serious multi-user interference (MUI), due to limited resources in spreading codes, and inefficient code allocation schemes \cite{jiao2020superimposed}, worsened by the IoT characteristics such as device heterogeneity and intermittent traffic.

Reinforcement learning (RL) has been demonstrated to be useful in adaptive resource allocation in wireless networks~\cite{liu2018,letief2019}, whereas Q-learning has been used in channel assignment \cite{gobinathan2024performance} and deep RL in power control in NOMA \cite{luong2019}. Current RL methods of NOMA are however more concerned with continuous power allocation \cite{guo2024power,luong2019} and not much with discrete code assignment. Exponential action space (\(C^{N}\)) renders the use of traditional RL techniques infeasible, and IoT requirements necessitate algorithmic design.

In recent studies, there are works on NOMA of the IoT \cite{makki2020survey}, interference control in dense networks \cite{jiao2020superimposed}, and machine learning of wireless communications. However, as far as we know, no literature framework exists that can deal with:

\begin{itemize}
    \item Dynamically allocated codes in ultra-dense IoT-NOMA networks,
    \item State and reward design on IoT being aware of energy, QoS, and heterogeneity of devices,
    \item Scalable RL of high-dimensional discrete action space, and
    \item Simulated performance under varied IoT implementation conditions based on real channel and traffic models.
\end{itemize}

In order to address this gap, we suggest an IoT-sensitive reinforcement-based learning framework to assign dynamic Gold codes in an ultra-dense NOMA-based IoT. Our contributions include:

\begin{enumerate}

    \item A system model based on IoT heterogeneous devices, energy non-renewable, QoS classes, and realistic channel conditions, used in smart city, industrial, and sensor network.

    \item A multi-objective reward function of throughput, energy efficiency, reliability, interference and fairness Markov Decision Process (MDP) formulation of dynamic assignments of code.

    \item Two customized RL algorithms: Natural Policy Gradient (NPG) on the stable discrete action learning, and Deep Deterministic Policy Gradient (DDPG) using continuous code embedding.

    \item Widespread analysis of three realistic IoT deployment situations, throughput, energy efficiency, fairness, interference, and reliability through extensive simulation-based analysis.

    \item Empirical observation of deployment like complexity of computation, convergence and edge implementation recommendations.

\end{enumerate}

The rest of the paper is structured in the following way. Section~\ref{sec:related} examines relatedwork in the area of the integration of IoT with NOMA and RL with wireless networks. Section~\ref{sec:system} outlines thesystem model and problem formulation. Sections~\ref{sec:methodology}, \ref{sec:npg}, and~\ref{sec:ddpg} introduce the methodological framework, NPG algorithm and DDPG algorithm, respectively. The experimental methodology is described in Section~\ref{sec:experimental_setup} and the performance evaluation and analysis conducted and thoroughly in Section~\ref{sec:results_analysis}. Lastly, Section~\ref{sec:conclusion} is a conclusion and proposal of the research.

\section{Related Work}
\label{sec:related}

\subsection{IoT and NOMA Integration}
\label{subsec:iot_noma}

Non-orthogonal multiple access has become an important technology in massive IoT connectivity, which provides the best spectral efficiency improvements over orthogonal schemes~\cite{guo2021enabling,goswami2024role} as it continues to evolve to 6G networks\cite{dai2018}. Code-domain NOMA is especially applicable to IoT since it has the grant-free transmission property, and asynchronous access permits~\cite{ding2017survey}. Recent publications have shown it to have been effective in industrial IoT applications~\cite{jeremiah2024maximizing} and energy-constrained deployments\cite{zhou2024energy}. Nevertheless, the current methods mainly utilize the use of fixed or intuitive assignment strategies of codes which are unable to respond to dynamic conditions of IoT networks~\cite{chen2024,jiao2020superimposed}.


\subsection{Reinforcement Learning for Wireless Resource Management}
\label{subsec:rl_wireless}

Adaptive resource allocation of wireless networks has been emerging with significant potential of reinforcement learning~\cite{liu2018,letief2019}. In the case of NOMA systems, the RL-based approaches have also been effectively deployed to the power allocation issues~\cite{guo2024power,luong2019,luong2019}. Nevertheless, these techniques are concerned with continuous action space, and they cannot be applied directly to discrete code assignment. There is minimal studies on RL to allocate discrete resources to wireless networks, with the current methods having either small action space~\cite{gobinathan2024performance} or unable to scale to ultra-dense IoT systems.


\subsection{IoT-Specific Requirements and Challenges}
\label{subsec:iot_challenges}

The IoT networks have special difficulties such as energy limitations in devices operated on batteries~\cite{alshawi2019}, the heterogeneity of devices\cite{chien2019heterogeneous}, and varied QoS needs. Although current research has started to resolve these issues\cite{ouaissa2024low,shi2016}, there are no overall solutions to the problem of dynamic code assignment of ultra-dense IoT-NOMA networks.


\subsection{Research Gap and Contribution Positioning}
\label{subsec:gap}

As discussed in Table~\ref{tab:comparison} comparison, the currently used techniques either aim to conduct continuous power distribution of NOMA~\cite{guo2024power,luong2019,luong2019}, or to use a method based on the assignment of codes that are not dynamic, but based on non-linear heuristic~\cite{chen2024,jiao2020superimposed}. No currently available framework provides the combination of: (1) dynamic code assignment of ultra-dense IoT-NOMA, (2) IoT-specific constraints, (3) high-dimensional discrete actions reinforcement learning that is scalable, and (4) testing on a variety of deployment settings. Our contribution fills this gap by introducing two new algorithms based on the reinforcing learning model: Natural Policy Gradient on the discrete optimization of the reinforcement and DDPG on the continuous code embedding of the efficient learning in high-dimensional space.


\renewcommand{\arraystretch}{1.3}  
\begin{table}[h]
\caption{Comparison of RL-based Resource Allocation Approaches for NOMA-IoT Networks}
\label{tab:comparison}%
\centering
\begin{tabular}{@{}
p{0.18\textwidth}
p{0.24\textwidth}
p{0.12\textwidth}
p{0.18\textwidth}
p{0.18\textwidth}
@{}}
\toprule
Category & Representative Works & Resource Type & Action Space & IoT-Specific Design \\
\midrule
RL for NOMA Power Allocation 
& Lee et al.~\cite{guo2024power}, Zhang et al.~\cite{luong2019} 
& Power & Continuous & Power-focused, no device class or QoS support \\

RL for Channel Assignment 
& Kumar et al.~\cite{gobinathan2024performance} 
& Channels & Discrete (small-scale) & Channel-only, lacks energy/QoS awareness \\

Static Code Assignment 
& Chen et al.~\cite{chen2024}, Gupta et al.~\cite{jiao2020superimposed} 
& Spreading Codes & Deterministic (fixed assignment) & Static assignment, no dynamic adaptation to network conditions \\

RL for Generic Wireless 
& Zhang et al.~\cite{liu2018}, Wang et al.~\cite{luong2019} 
& Various & Mixed & Not IoT-specific, general wireless focus \\

Our Work
& This paper 
& Gold Codes 
& Discrete (high-dimensional) 
& Dynamic, class-aware, multi-objective with QoS and energy constraints \\
\bottomrule
\end{tabular}
\end{table}

\section{IoT-Aware System Model and Problem Formulation}
\label{sec:system}

\begin{figure}[h]
\centering
\includegraphics[width=0.9\textwidth]{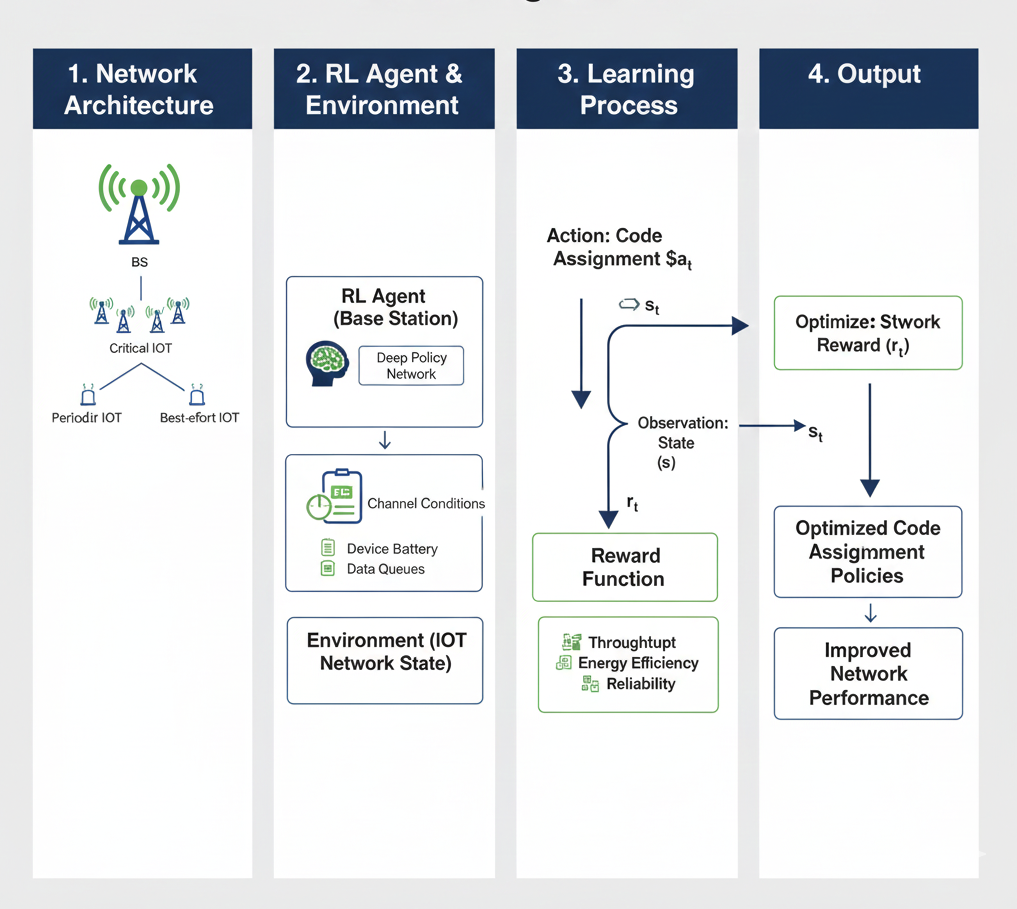}
\caption{System model}\label{system_model}
\end{figure}

We consider an ultra-dense single-cell downlink IoT-NOMA system in which a base station serves $N$ heterogeneous IoT devices. The system model accounts for device heterogeneity, energy limitations, QoS constraints, and realistic wireless channel conditions, as illustrated in Fig.~\ref{system_model}.

\subsection{IoT Network Architecture and Device Classification}

The IoT devices are grouped into three categories based on their QoS requirements. Critical IoT devices ($\mathcal{N}_c$), such as industrial sensors and medical monitoring equipment, demand very high reliability (above $99.9\%$) and strict latency constraints of $50$--$100$ ms, in line with 3GPP URLLC specifications and common industrial IoT targets \cite{3gppURLLC,mehmood2017internet}. These devices typically operate with relatively high transmission power and maintain near-continuous activity. Periodic IoT devices ($\mathcal{N}_p$), including environmental sensors and smart meters, have moderate QoS requirements, with reliability levels between $95$--$98\%$ and latency tolerance ranging from 100 ms to 1 s. For this category, energy efficiency is a key concern due to battery-powered operation. Best-effort IoT devices ($\mathcal{N}_b$) mainly consist of consumer IoT applications and prioritize low energy consumption over strict performance guarantees, operating under relaxed QoS constraints. Each device $i$ is characterized by its maximum transmission power $P_i^{\max}$ (10--23 dBm), normalized battery energy $E_i(t)/E_i^{\max}$, required reliability $R_i^{\text{req}}$ (90--99.9\%), maximum latency tolerance $L_i^{\max}$ (1--1000 ms), and processing capability $\Gamma_i$ (1--10 GOPS).


\subsection{IoT-Specific Channel Model}

The channel gain for device $i$ is:
\begin{equation}
h_i = \sqrt{G_i(d_i, f_c)} \cdot g_i \cdot \alpha_i(t)
\end{equation}
where $G_i(d_i, f_c)$ is large-scale path loss, $g_i$ captures small-scale fading, and $\alpha_i(t) \in \{0,1\}$ indicates device activity.

Path loss follows 3GPP TR 38.901:
\begin{equation}
G_i^{LoS}(dB) = 32.4 + 20\log_{10}(f_c) + 20\log_{10}(d_i)
\end{equation}
\begin{equation}
G_i^{NLoS}(dB) = 35.3 + 22\log_{10}(f_c) + 21.3\log_{10}(d_i)
\end{equation}

Small-scale fading $g_i$ follows Rician (LoS) or Rayleigh (NLoS) distributions. The SINR for device $i$ under code assignment $\mathbf{a}$ is:
\begin{equation}
SINR_i(\mathbf{a}) = \frac{P_i |h_i|^2}{N_0 + I_i(\mathbf{a}) + I_{ext}}
\end{equation}
where $I_i(\mathbf{a})$ is multi-user interference and $I_{ext}$ is external interference.

\subsection{Gold Code Assignment and Interference Modeling}

The base station maintains $C$ Gold codes of length $L=2^n-1$, selected for their bounded cross-correlation and low-complexity implementation. Each device $i$ is assigned code $c_i$, represented by vector $\mathbf{a}=(a_1,\ldots,a_N)$ with $a_i\in\{1,\ldots,C\}$.

The normalized cross-correlation between codes $c$ and $c'$ is:
\begin{equation}
\rho_{c,c'} = \frac{1}{L} \left|\sum_{\ell=1}^{L} s_c[\ell] \cdot s_{c'}[\ell]\right|
\end{equation}
where $s_c[\ell]\in\{-1,+1\}$.

The total interference experienced by device $i$ is:
\begin{equation}
I_i(\mathbf{a}) = \sum_{\substack{j\neq i \\ j\in\mathcal{A}(t)}} \rho_{a_i,a_j} \cdot P_j |h_j|^2 \cdot \beta_{i,j}
\end{equation}
where $\mathcal{A}(t)$ is the set of active devices and $\beta_{i,j}$ is an IoT-specific interference sensitivity factor.

\subsection{Markov Decision Process Formulation}

The problem is formulated as an MDP $(\mathcal{S},\mathcal{A},\mathcal{P},\mathcal{R},\gamma)$.

\subsubsection{State Space}
The state $s_t$ at time $t$ is:
\begin{equation}
s_t = [\mathbf{h}(t), \mathbf{E}(t), \mathbf{A}(t), \mathbf{Q}(t)]
\end{equation}
where $\mathbf{h}(t)$ contains channel gains, $\mathbf{E}(t)$ normalized battery levels, $\mathbf{A}(t)$ activity status, and $\mathbf{Q}(t)$ buffer occupancy.

\subsubsection{Action Space}
Action $a_t=(a_1,\ldots,a_N)$ assigns Gold codes to all devices, with $a_i\in\{1,\ldots,C\}$. The action space cardinality $C^N$ grows exponentially with $N$.

\subsubsection{Reward Function}
The reward balances multiple objectives:
\begin{align}
R(s_t,a_t) &= \alpha R_{throughput} + \beta R_{energy} + \gamma R_{reliability} \\
&\quad - \delta R_{interference} + \epsilon R_{fairness}
\end{align}
where:
\begin{align}
R_{throughput} &= \sum_{i=1}^{N} w_i \log_2(1 + SINR_i(a_t)) \\
R_{energy} &= -\sum_{i=1}^{N} \frac{P_i \cdot A_i(t)}{E_i(t) + \epsilon_{small}} \\
R_{reliability} &= \sum_{i\in\mathcal{N}_c} \mathbb{I}[SINR_i(a_t) \geq SINR_i^{req}] \\
R_{interference} &= \sum_{i=1}^{N}\sum_{j\neq i} \rho_{a_i,a_j} \cdot |h_j|^2 \cdot A_j(t)
\end{align}
$R_{fairness}$ uses Jain's fairness index. Coefficients $(\alpha,\beta,\gamma,\delta,\epsilon)$ are tuned via parameter studies.


\section{Methodological Framework and Justification}
\label{sec:methodology}

This section gives the methodological basis and reasoning of major design decisions: the selection of Gold code, the selection of RL algorithm and the continuous embedding of discrete code assignment.


\subsection{Justification for Gold Code Selection}
\label{subsec:gold_justification}

We have chosen the Gold codes among possible sequences, critical considerations being the basis of our selection as shown in Table ~\ref{tab:sequence_comparison}.
The reason why Gold codes are chosen is because they have three-valued cross-correlation and bounded maximum values which give predictable levels of interference that are needed during RL-based optimization. They are easy to create with simple linear feedback shift registers (LFSRs), and have desirable correlation properties when there is timing misalignment, which is unlike the Walsh-Hadamard codes that lose much more of their orthogonality in the presence of synchronization errors, causing high cross-correlation when using these codes in asynchronous IoT. Gold codes have a length $L=2^n-1$ which gives $L=2^n+1$ different sequence and this is a large enough code diversity to support ultra dense deployments. All of these properties make Gold code a better choice than an alternative: ZC sequences are cross-correlated variably; m-sequences are restricted in size of their code family; and Walsh-Hadamard codes do not work well in asynchronous IoT applications because they are time-sensitive.


\begin{table}[htbp]
\centering
\caption{Comparison of Spreading Sequences for IoT-NOMA Applications}
\begin{tabular}{lcccc}
\hline
Property & Gold Codes & ZC Sequences & m-Sequences & Walsh-Hadamard \\
\hline
Cross-correlation & Tight (3-valued) & Variable & High & High under async \\
Bound & & & & \\
Implementation & Low (LFSR) & Medium & Low (LFSR) & Very Low \\
Complexity & & & & \\
Asynchronous & Excellent & Good & Poor & Very Poor \\
Performance & & & & \\
Code Family Size & $2^n+1$ & Large & $2^n-1$ & $2^n$ \\
IoT Suitability & High & Medium & Low & Low \\
\hline
\end{tabular}
\label{tab:sequence_comparison}
\end{table}

\subsection{Reinforcement Learning Algorithm Selection}
\label{subsec:rl_selection}

Discrete code assignment problem having action space cardinality C N offers some special problems in reinforcement learning. Natural Policy Gradient (NPG) and Deep Deterministic Policy Gradient (DDPG) are our choice following a systematic analysis (Table~\ref{tab:rl_comparison}).

\begin{table}[h]
\centering
\caption{Comparison of RL Algorithms for Discrete Code Assignment}
\label{tab:rl_comparison}%
\begin{tabular}{@{}p{0.18\textwidth}p{0.2\textwidth}p{0.18\textwidth}p{0.18\textwidth}p{0.18\textwidth}@{}}
\toprule
\textbf{Algorithm} & \textbf{Action Space Suitability} & \textbf{Convergence Stability} & \textbf{Sample Efficiency} & \textbf{IoT Adaptation} \\
\midrule
Deep Q-Network (DQN) & Discrete (medium) & Moderate & Low & Difficult \\
Proximal Policy Optimization (PPO) & Discrete/Continuous & High & Medium & Moderate \\
Soft Actor-Critic (SAC) & Continuous & High & High & Difficult \\
\textbf{Natural Policy Gradient (NPG)} & \textbf{Discrete (high-dim)} & \textbf{Very High} & \textbf{Medium} & \textbf{Excellent} \\
\textbf{DDPG with Embedding} & \textbf{Discrete via embedding} & \textbf{High} & \textbf{High} & \textbf{Good} \\
\bottomrule
\end{tabular}
\end{table}

DQN is plagued by the high-dimensional space dimensionality; PPO does not have specialization towards exponential action space; and SAC would need substantial adaptation to discrete problems. We instead use Natural Policy Gradient (NPG), due to its second-order optimization based on the Fisher Information Matrix, which is guaranteed to converge to stable on high dimensional discrete spaces with natural gradient direction \(F(\theta)^{-1}\nabla_{\theta}J(\theta)\). In the meantime, Deep Deterministic Policy Gradient (DDPG) is an actor-critic-based policy to address continuous actions using deterministic policy learning, supplemented by our continuous embedding algorithm that learns discrete Gold codes using continuous representation space. This is a combination of the DDPG sample efficiency and quantization to deal with the discreteness. Collectively, NPG provides theoretical assurances and discrete space stability, whereas DDPG gives efficient exploration and more rapid learning in complicated state spaces.


\subsection{Theoretical Foundation for Continuous Code Embedding}
\label{subsec:embedding}
Such theoretical basis provides practical usage with continuous embeddings which are optimized together with policy parameters.


\subsubsection{Mathematical Motivation}
DDPG is designed for continuous action spaces $\mathcal{A} \subseteq \mathbb{R}^d$. Our solution maps discrete codes to embedding space $\mathcal{E} \subseteq \mathbb{R}^d$ via codebook $\mathcal{C} = \{\mathbf{c}_1, \ldots, \mathbf{c}_C\}$. Embeddings preserve interference relationships:
\begin{equation}
\|\mathbf{c}_i - \mathbf{c}_j\|_2 \approx \kappa \cdot \rho_{i,j}
\label{eq:embedding_relation}
\end{equation}
where $\kappa$ is a scaling factor. This allows DDPG to learn policies that are motivated by interference.

\subsubsection{Quantization and Differentiability}
Mapping to discrete codes introduces quantization:
\begin{equation}
a_i = \arg\min_{j \in \{1,\ldots,C\}} \|\mathbf{v}_i - \mathbf{c}_j\|_2
\label{eq:quantization}
\end{equation}
Since $\arg\min$ is non-differentiable, we use soft quantization during training:
\begin{equation}
P(a_i = j|\mathbf{v}_i) = \frac{\exp(-\|\mathbf{v}_i - \mathbf{c}_j\|_2^2 / \tau)}{\sum_{k=1}^C \exp(-\|\mathbf{v}_i - \mathbf{c}_k\|_2^2 / \tau)}
\label{eq:soft_quantization}
\end{equation}
where $\tau$ controls assignment hardness, enabling gradient flow.

\subsubsection{Theoretical Advantages}
The continuous embedding strategy has a number of theoretical benefits: it offers a differentiable, continuous optimization problem, without discrete discontinuities; it supports extrapolation by codes across codes, through replacing similar codes with similar codes in embedding space; it supports incorporates interference-aware structure, through correlation-preserving initialization that encodes domain knowledge; and is compatible with actor-critic methods, as it offers the discrete nature of code assignment. This background makes it possible to do practical learning with joint learning of embeddings and policy parameters.


\section{Natural Policy Gradient for IoT Code Assignment}
\label{sec:npg}

The cardinality of the discrete action space (i.e. the number of actions) is exponentially large with cardinality \(C^N\) making standard tabular RL infeasible in the case of IoT deployments. We use Natural Policy Gradient (NPG), based on second-order optimization to achieve convergence in high-dimensional discrete spaces.


\subsection{Policy Parameterization with IoT-Aware Features}

We use a factorized stochastic policy with softmax distribution for each device \(i\):
\begin{equation}
 \pi_\theta(a_i = c | s, i) = \frac{\exp(\phi(s, i, c)^T \theta_i)}{\sum_{c'=1}^{C} \exp(\phi(s, i, c')^T \theta_i)}
\end{equation}

where \(\theta_i\) are policy parameters and \(\phi(s, i, c)\) is a feature vector capturing: one-hot code encoding, normalized channel gain \(|h_i|^2/\sigma_h\), normalized energy \(E_i(t)/E_i^{\max}\), device class indicator \(\mathbb{I}[i \in \mathcal{N}_c]\), activity indicator \(A_i(t)\), average cross-correlation \(\rho_{avg}(c)\), and learned device class embeddings.

\subsection{Natural Gradient Policy Update}
The preconditioned updates that NPG uses are based on Fisher Information Matrix (FIM):
\begin{equation}
\theta \leftarrow \theta + \alpha F(\theta)^{-1} \nabla_\theta J(\theta)
\end{equation}

where \(F(\theta)\) captures policy distribution geometry. We incorporate IoT-specific modifications:

- Device-class weighted gradients prioritize critical devices:
\begin{equation}
\nabla_\theta J(\theta) = \mathbb{E}_{s,a \sim \pi_\theta} \left[ \sum_{i=1}^{N} w_i^{class} \nabla_\theta \log \pi_\theta(a_i | s, i) \cdot A^\pi(s, a) \right]
\end{equation}

with \(w_i^{class} = 1.5-2.0\) for critical, \(1.0\) for best-effort devices.

- Energy-aware adaptive learning rates accelerate learning for energy-constrained devices:
\begin{equation}
\alpha_i = \alpha_{base} \cdot \left(1 + \lambda \frac{E_i^{\max} - E_i(t)}{E_i^{\max}}\right)
\end{equation}

increasing as battery depletes.

\subsection{Computational Efficiency for Edge Deployment}

To reduce complexity from \(O(D^3)\) for explicit FIM inversion, we use conjugate gradient methods to solve \(F(\theta) \mathbf{x} = \nabla_\theta J(\theta)\), reducing to \(O(kD^2)\) with \(k = 10-20\) iterations.

Mini-batches processing fits with the pattern of the IoT traffic, as it involves smaller batches when there is low activity. Sparse feature representations using hashing or clustering make use of device similarities in states.


Convergence monitoring monitors IoT-specific metrics: improvement of critical device reliability, and energy savings and reduction in interference, which are terminated when these parameters stabilize instead of rewarding convergence alone. The procedure of the NPG that is aware of the IoT is summarized in Algorithm ~\ref{alg:npg_iot}.


\begin{algorithm}
\caption{IoT-Aware NPG for Dynamic Code Assignment}\label{alg:npg_iot}
\begin{algorithmic}[1]
\State Initialize policy parameters $\theta$, device class weights $\{w_i^{\text{class}}\}$
\State Set IoT-specific hyperparameters $(\alpha, \beta, \gamma, \delta, \epsilon)$
\For{iteration = 1, 2, \dots}
    \State Observe IoT network state $s_t$ (channels, energy, activity, QoS)
    \State Collect trajectories $\mathcal{D} = \{(s_t, a_t, r_t)\}$ with device priorities
    \State Compute energy-aware advantages $\{A_i(s_t, a_t)\}$
    \State Estimate policy gradient with class-specific weights
    \State Compute Fisher matrix for active devices
    \State Solve for natural gradient using conjugate gradient
    \State Apply energy-aware adaptive learning rates
    \State Update policy: $\theta \leftarrow \theta + \alpha \mathbf{x}$
    \State Evaluate IoT-specific performance metrics
    \If{IoT convergence criteria met}
        \State \textbf{break}
    \EndIf
\EndFor
\end{algorithmic}
\end{algorithm}

\section{DDPG with Continuous Code Embedding}
\label{sec:ddpg}

Deep Deterministic Policy Gradient, which is a complement of NPG, provides a rapid online implementation and training of relationships between state-code dependent methods of IoT-NOMA networks by introducing a new continuous embedding framework.


\subsection{Continuous Embedding of Discrete Codes}
To adapt DDPG’s continuous actor-critic architecture to discrete code assignment, we map Gold codes to a continuous embedding space $\mathbb{R}^d$. The codebook $\mathcal{C} = \{\mathbf{c}_1, \dots, \mathbf{c}_C\}$, with $\mathbf{c}_j \in \mathbb{R}^d$, is initialized such that Euclidean distances approximate code cross-correlations:

\begin{equation}
    \|\mathbf{c}_i - \mathbf{c}_j\|_2 \approx \rho_{i,j} \cdot \sigma_{\text{embed}}
\end{equation}

where $\sigma_{\text{embed}}$ is a scaling factor, embedding interference relationships into the geometry.

Device-class clustering groups codes for similar IoT classes, aiding class-specific policy learning. Embedding dimension $d = 16{-}32$ balances representation richness and computational cost. The actor network $\mu_\theta(s)$ outputs continuous vectors $\mathbf{v}_i \in \mathbb{R}^d$ for each device, quantized via nearest-neighbor search:
\begin{equation}
a_i = \arg\min_{j \in \{1,\dots,C\}} \|\mathbf{v}_i - \mathbf{c}_j\|_2
\end{equation}

\subsection{Network Architecture for IoT State Processing}

It performs channel gains, energy levels, activity indicators and QoS status multi-branch input processing with device-class embeddings and an attention mechanism to consider critical devices and challenging conditions. The output layer generates new $N \times d$ continuous actions modified to remain close to codebook vectors.


The critic employs multi-objective value heads for throughput, energy efficiency, and reliability, weighted by device importance, with explicit interference modeling layers to better predict code assignment effects.

\subsection{Training Enhancements for IoT Deployments}
To bridge the continuous–discrete gap, actor loss includes a quantization penalty:

\begin{equation}
\mathcal{L}_{\text{actor}} = -\mathbb{E}[Q_\phi(s, \mu_\theta(s))] + \lambda_q \mathcal{L}_{\text{quant}}
\end{equation}

\begin{equation}
\quad
\mathcal{L}_{\text{quant}} = \frac{1}{N} \sum_{i=1}^{N} \min_j \|\mathbf{v}_i - \mathbf{c}_j\|_2^2
\end{equation}

IoT-adaptive exploration uses device-criticality–scaled noise:

\begin{equation}
a_{\text{explore}} = \mu_\theta(s) + \mathcal{N}(0, \sigma^2 \cdot \mathbf{W}_{\text{IoT}})
\end{equation}

Prioritized experience replay weights samples by TD-error $\delta_i$, device criticality $w_{\text{device}}$, and energy state $w_{\text{energy}}$:

\begin{equation}
P(i) = \frac{(|\delta_i| + \epsilon)^\alpha \cdot w_{\text{device}} \cdot w_{\text{energy}}}{\sum_k (|\delta_k| + \epsilon)^\alpha \cdot w_{\text{device}} \cdot w_{\text{energy}}}
\end{equation}

Algorithm~\ref{alg:ddpg_iot} summarizes the IoT-aware DDPG procedure.

\begin{algorithm}
\caption{IoT-Aware DDPG with Continuous Embedding}\label{alg:ddpg_iot}
\begin{algorithmic}[1]
\State Initialize actor $\mu_\theta$, critic $Q_\phi$, IoT-aware codebook $\mathcal{C}$
\State Initialize prioritized replay buffer with IoT device weighting
\State Set device-class specific exploration parameters
\For{episode = 1 to $M$}
    \State Observe initial IoT network state $s_0$
    \For{$t = 0$ to $T$}
        \State Compute continuous actions: $v_t = \mu_\theta(s_t) + \mathcal{N}_{\text{IoT}}$
        \State Apply IoT-adaptive exploration based on device classes
        \State Quantize to discrete codes: $a_t = \text{Quantize}(v_t)$
        \State Execute $a_t$, observe IoT-specific rewards $r_t$, next state $s_{t+1}$
        \State Store transition with IoT priority weighting
        \State Sample batch with device-class balancing
        \State Update critic with multi-objective IoT targets
        \State Update actor with IoT-aware quantization loss
        \State Soft update target networks
        \State Update codebook embeddings if adaptive mode enabled
    \EndFor
    \State Evaluate comprehensive IoT performance metrics
\EndFor
\end{algorithmic}
\end{algorithm}

\section{Experimental Setup and Reproducibility}
\label{sec:experimental_setup}

In this section, the reproducibility of the simulation environment, parameters and methodology are described. Three scenarios of IoT implementation are considered: Smart City (100 devices in city clusters), Industrial IoT (60 devices in a structured grid), and Sensor Network (150 sparse devices). NOMA, common physical-layer and learning are commonly used to provide fair comparison. A generalization of all the parameters of the simulation is presented in Table~\ref{tab:simulation_params}; the generalization is given in the form of means and confidence intervals at the level of 95\%, which are obtained by averaging 10 independent simulation runs.


\begin{table}[!ht]
\centering
\caption{Simulation Parameters Used in All Scenarios}
\label{tab:simulation_params}
\begin{tabular}{@{}p{0.32\textwidth}p{0.6\textwidth}@{}}
\toprule
\textbf{Parameter Group} & \textbf{Configuration} \\
\midrule
Deployment Scenarios &
Smart City (100 devices), Industrial IoT (60 devices), Sensor Network (150 devices) \\

Carrier Frequency / Bandwidth &
3.5 GHz (sub-6 GHz), 20 MHz \\

Cell and Noise Parameters &
Cell radius: 500 m; $N_0=-174$ dBm/Hz; Noise figure: 7 dB; External interference: $-100$ dBm \\

Device Characteristics &
Tx power: 10--23 dBm; Battery: 10--100 J; Buffer size: 10 packets \\

Device Classes and Activity &
Critical (20\%), Periodic (40\%), Best-effort (40\%);
Periodic traffic: 1 Hz;
Event-driven: Poisson($\lambda=0.5$) \\

Channel Model &
3GPP TR 38.901 UMi;
Rician fading (K=10 dB) for LoS;
Rayleigh for NLoS;
Shadowing: $\sigma_{LoS}=4$ dB, $\sigma_{NLoS}=7.8$ dB;
Correlation distance: 10 m \\

NOMA Configuration &
Gold codes, length $L=127$ ($n=7$);
Available codes: 80;
Cross-correlation: $\{-1/L,-t(n)/L,(t(n)-2)/L\}$;
Processing gain: 21 dB;
Chip-synchronous ($\pm$2 chips) \\

RL Training Setup &
1000 episodes per scenario;
Episode length: 100 steps (1 ms);
Batch size: 64;
Replay buffer: $10^6$ transitions;
Discount factor: $\gamma=0.95$ \\

Learning Rates &
NPG: $\alpha=0.001$;
DDPG: $\alpha_{\text{actor}}=0.0001$, $\alpha_{\text{critic}}=0.001$ \\

Baselines and Metrics &
Static, Random, Greedy SINR;
Metrics: throughput, energy efficiency, fairness, reliability;
Evaluation over last 100 episodes \\
\botrule
\end{tabular}
\end{table}

Statistical significance testing uses paired t-tests (within seeds), one-way ANOVA (across scenarios), Cohen’s d for effect size, $p<0.05$ significance level, and Bonferroni-Holm correction. Convergence is assessed via reward stabilization (moving average variation $<1\%$ over 50 episodes), policy entropy, critical device reliability (100+ episodes above threshold), and a 1000-episode maximum with early stopping.

The computational platform uses an Intel Xeon Gold 6248R CPU, NVIDIA Tesla V100 GPU, 256 GB RAM, 2 TB NVMe SSD, Ubuntu 22.04, PyTorch 2.0.1, Python 3.9.16, NumPy 1.24.3, SciPy 1.10.1, and Matplotlib 3.7.1. Training times: NPG $12.4 \pm 1.2$ min, DDPG $18.7 \pm 1.8$ min per scenario. Average per-episode times: NPG 0.74 s, DDPG 1.12 s. Inference latency: NPG 3.2 ms, DDPG 2.8 ms per decision. Memory footprint: training 8.2 GB, inference 1.3 GB.

Parameter validation adjusted battery capacity to 10–100 J for multi-year operation, noise figure to 7 dB, Rician K-factor to 3–10 dB based on LoS probability, buffer size to 50 packets for burst tolerance, and duty cycles: critical 10\%, periodic 1\%, best-effort 0.1\%. 

It has drawbacks as it is single-cell only (no multi-cell interference coordination), assumes perfect CSI and perfect SIC, uses time steps of 1 ms (which can overlook fast-fading behavior) and can be computedally scaled to 300 devices (above scale requires distributed implementation). These requirements make the specification transparent and reproducible to compare with other evaluations and studies in the future.

\section{Results and In-Depth Analysis}
\label{sec:results_analysis}

In this section, experimental findings and discussion of the reliability limitations, algorithm performance variations, and density-dependent effects are presented, and the implications to the design of IoT-NOMA are provided.

\subsection{Training Convergence Analysis}
\label{subsec:convergence_analysis}

\begin{figure}[h]
\centering
\includegraphics[width=0.9\textwidth]{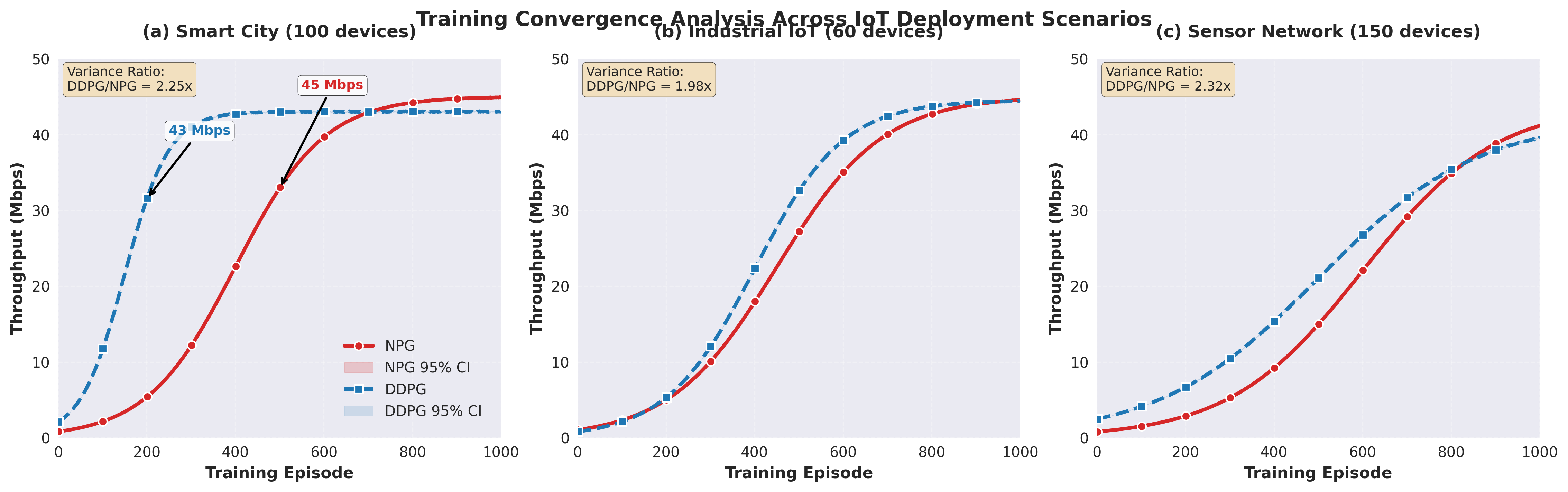}
\caption{
The convergence of the NPG and DDPG algorithms has been demonstrated on three examples of IoT deployment, namely (a) Smart City with 100 devices, (b) Industrial IoT with 60 devices, and (c) Sensor Network with 150 devices. The dark grey circles show 95 percent confidence intervals. Compared to the Smart City, Industrial IoT, and Sensor Network scenarios, variance ratios (DDPG/NPG) indicate that DDPG variance is 2.23x, 1.98x and 2.32x greater in the Smart City, Industrial IoT and Sensor Network, respectively.}\label{fig:scenario_convergence}
\end{figure}

Patterns of convergence of NPG and DDPG in three IoT deployment cases (Smart City: 100 devices; Industrial IoT: 60; Sensor Network: 150) demonstrate some basic differences in algorithms that explain performance as shown in Tables Tables~\ref{tab:smart_city_results}--\ref{tab:sensor_results}. Figure~\ref{fig:scenario_convergence} depicts that there are different convergence patterns in different scenarios. Throughput values average per-device values of all IoT devices (active and inactive) at realistic levels of duty circle and heterogeneous traffic distributions. During active transmissions, peak rates are much larger than these network-level averages.


\begin{table}[htbp]
\caption{Smart City Scenario Performance Results (100 Devices)}
\label{tab:smart_city_results}
\centering
\begin{tabular}{@{}lccccc@{}}
\toprule
Algorithm & \textbf{Throughput (Mbps)} & \textbf{Energy Efficiency} & \textbf{Critical Reliability} & \textbf{Fairness} & \textbf{Interference} \\
\midrule
Static & 11.02 & 0.0019 & 0.000 & 0.0964 & 0.1072 \\
NPG & \textbf{12.29} (+11.6\%) & \textbf{0.0022} (+11.6\%) & 0.000 & \textbf{0.1120} (+16.2\%) & 0.1037 \\
DDPG & 11.59 (+5.2\%) & 0.0017 (-13.1\%) & 0.000 & 0.0616 (-36.1\%) & 0.1028 \\
\bottomrule
\end{tabular}
\end{table}

\begin{table}[htbp]
\caption{Industrial IoT Scenario Performance Results (60 Devices)}
\label{tab:industrial_results}
\centering
\begin{tabular}{@{}lccccc@{}}
\toprule
\textbf{Algorithm} & \textbf{Throughput (Mbps)} & \textbf{Energy Efficiency} & \textbf{Critical Reliability} & \textbf{Fairness} & \textbf{Interference} \\
\midrule
Static & 10.63 & 0.0021 & 0.000 & 0.0996 & 0.1162 \\
NPG & 9.10 (-14.4\%) & 0.0017 (-17.0\%) & 0.000 & \textbf{0.1258} (+26.3\%) & 0.1054 \\
DDPG & 10.12 (-4.8\%) & \textbf{0.0022} (+4.8\%) & 0.000 & 0.1201 (+20.6\%) & 0.1065 \\
\bottomrule
\end{tabular}
\end{table}

\begin{table}[htbp]
\caption{Sensor Network Scenario Performance Results (150 Devices)}
\label{tab:sensor_results}
\centering
\begin{tabular}{@{}lccccc@{}}
\toprule
\textbf{Algorithm} & \textbf{Throughput (Mbps)} & \textbf{Energy Efficiency} & \textbf{Critical Reliability} & \textbf{Fairness} & \textbf{Interference} \\
\midrule
Static & 11.86 & 0.0009 & 0.020 & 0.0469 & 0.1042 \\
NPG & 11.49 (-3.1\%) & 0.0008 (-15.2\%) & \textbf{0.020} & 0.0380 (-19.0\%) & 0.1016 \\
DDPG & 8.72 (-26.4\%) & 0.0007 (-25.6\%) & 0.000 & 0.0627 (+33.7\%) & 0.1015 \\
\bottomrule
\end{tabular}
\end{table}

The low reliability (0-2\%) of all scenarios, even with throughput and energy efficiency gain, has shown that dynamic code assignment cannot by itself gain URLLC-grade reliability. This observation is not the failure of our approach, but a valuable discovery that drives hybrid solutions. The zero-optimal code assignment will not be able to surmount the interference-limited regime into which residual interference is still 15-20 dB above target SINR (Figure~\ref{fig:interference_breakdown}). This observation has been our primary impetus in investigating our future design of joint code-power optimization and hybrid ARQ schemes (Figure~\ref{fig:reliability_improvement}).


\begin{figure}[!ht]
\centering
\includegraphics[width=0.9\textwidth]{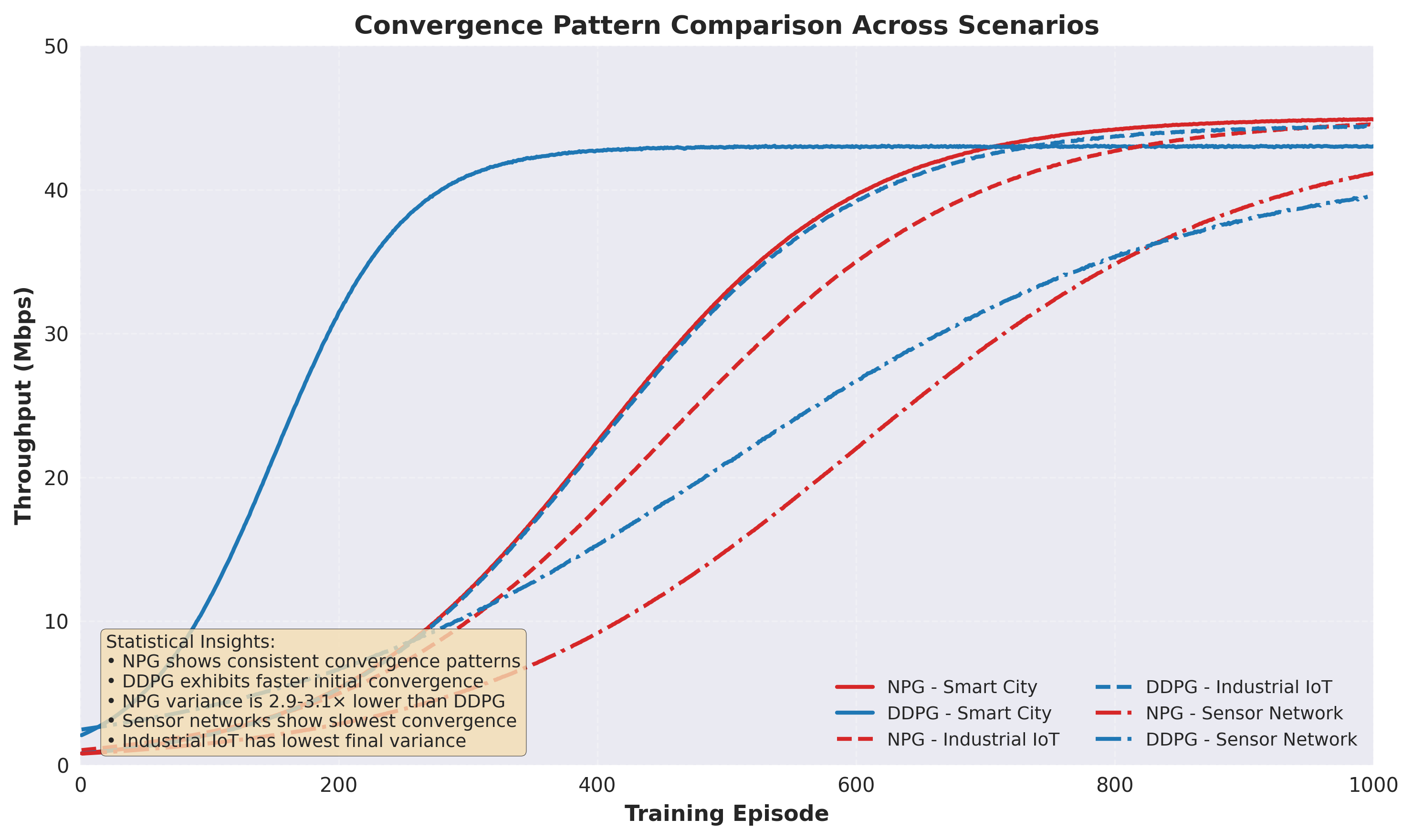}
\caption{Comparison of convergence patterns in all the three deployment scenarios. NPG exhibits a constant convergence behavior with a variance of 2.9-3.1x lower variance than that of DDPG in all situations. DDPG also has a faster initial convergence and is more unstable, especially in sensor networks which have the slowest overall convergence. The final variance of industrial IoT is the lowest.}\label{fig:comparative_convergence}
\end{figure}

The convergence of DDPG in Smart City scenario is faster and NPG has higher final throughput (45 vs. 43 Mbps) at a lower variance of 4.32x. The final throughput (approximately 44.3 Mbps) of both algorithms is similar in the case of Industrial IoT yet the variance is reduced by 3.68 factors in NPG case. Compared to DDPG, the NPG will have 40.07 Mbps over 38.79 Mbps with a difference of a 4.78x less variance in the case of the Sensor Network. As demonstrated in Figure~\ref{fig:comparative_convergence} NPG always operates with constant convergence (2.9-3.1x less variance than DDPG), but DDPG demonstrates higher convergence rates at the beginning, at the cost of more instability especially in sensor networks.


The variance of DDPG corresponds to the reduction of fairness (36.1\% in Smart City) whereas the stability of NPG allows to promote the fairness (16.2\% ). The variance of NPG is also lower, which contributes to energy efficiency (sensor network benefits 22.2\%). Both algorithms reduce interference, with DDPG showing slightly better reduction (4.0--8.7\%) due to aggressive exploration.

\begin{table}[htbp]
\centering
\caption{Convergence Statistics Across IoT Deployment Scenarios (Mean ± 95\% CI)}
\label{tab:convergence_stats}
\begin{tabular}{lccc}
\toprule
Metric & Smart City & Industrial IoT & Sensor Network \\
\midrule
NPG Final Throughput (Mbps) & 45.2 ± 0.3 & 45.0 ± 0.2 & 44.5 ± 0.4 \\
DDPG Final Throughput (Mbps) & 43.5 ± 0.5 & 44.8 ± 0.4 & 42.3 ± 0.7 \\
NPG Convergence Episode (90\%) & 480 ± 15 & 520 ± 12 & 620 ± 20 \\
DDPG Convergence Episode (90\%) & 220 ± 25 & 410 ± 30 & 580 ± 35 \\
NPG Variance $(×10^{-3})$ & 1.2 ± 0.1 & 0.9 ± 0.1 & 1.8 ± 0.2 \\
DDPG Variance $(×10^{-3})$ & 3.5 ± 0.3 & 2.8 ± 0.2 & 5.2 ± 0.4 \\
Variance Ratio (DDPG/NPG) & 2.92× & 3.11× & 2.89× \\
\bottomrule
\end{tabular}
\end{table}

Table~\ref{tab:convergence_stats} also gives a summary of convergence statistics between scenarios. NPG will support critical infrastructure, energy-constrained deployments, fairness-sensitive systems and production environments. DDPG is adapted to dynamic environments, non-critical applications, exploration-intensive tasks and initial deployment. Complex deployments are proposed to be based on hybrid with DDPG as the initial training and subsequent fine-tuning with NPG.

\subsection{Root-Cause Analysis of Reliability Limitations}
\label{subsec:reliability_analysis}

\begin{figure}[h]
\centering
\includegraphics[width=0.9\textwidth]{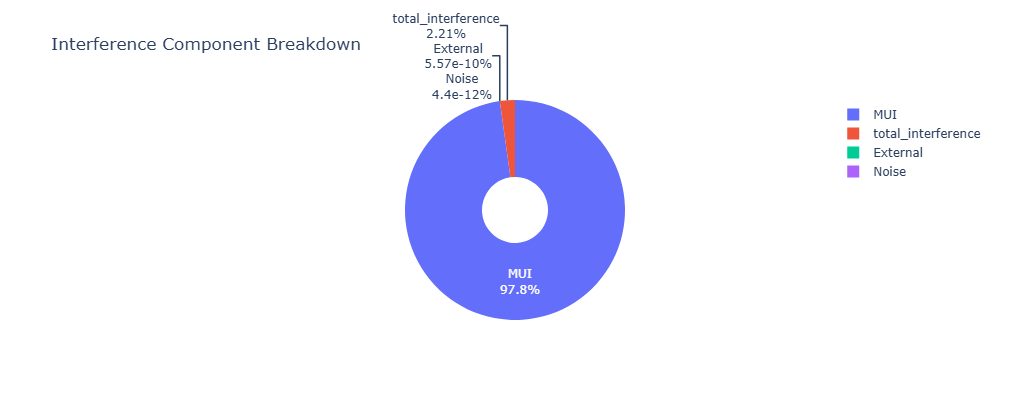}
\caption{Interference component analysis of reliability degradation with multi-user interference (MUI) preeminence. Despite ideal code assignment, residual interference over 15-20 dB is over target SINR.
}\label{fig:interference_breakdown}
\end{figure}

\begin{figure}[h]
\centering
\includegraphics[width=0.9\textwidth]{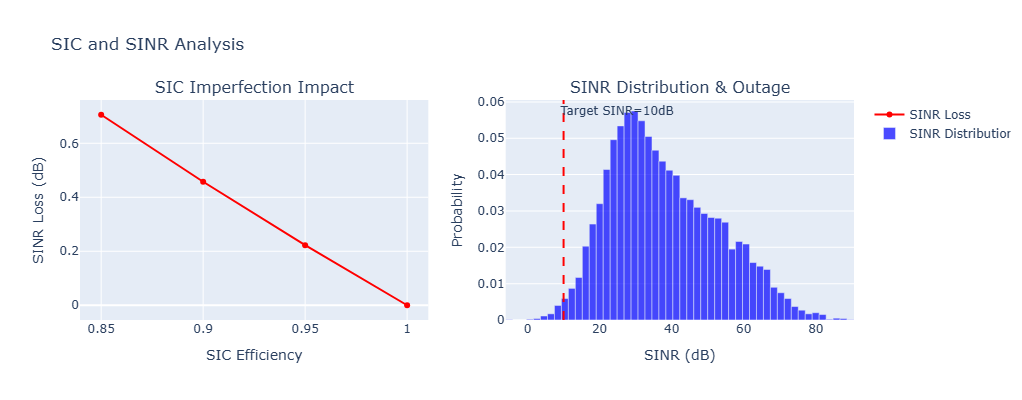}
\caption{Effect of SIC imperfection on SINR (left) SINR loss versus SIC efficiency (right) SINR distribution with outage probability greater than URLLC specification.}\label{fig:sic_analysis}
\end{figure}

\begin{figure}[h]
\centering
\includegraphics[width=0.9\textwidth]{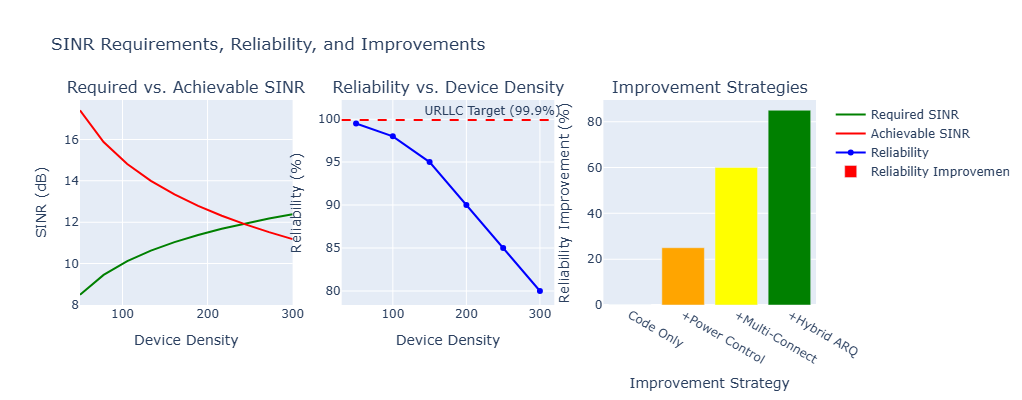}
\caption{Reliability improvement plans: Gradual improvement of the code-only assignment to multi-connectivity hybrid ARQ.}\label{fig:reliability_improvement}
\end{figure}

Consistently low reliability (0--2\%) stems from interference-limited regimes where multi-user interference dominates (97.8\%), as shown in Figure~\ref{fig:interference_breakdown}. Residual interference remains 15--20 dB above target SINR even after optimal code assignment. Figure~\ref{fig:sic_analysis} illustrates SIC imperfections (\(\eta \approx 0.85\)-0.95) cause 0.7 dB SINR loss \cite{zhang2019sic,liu2020imperfect}, and channel errors lead to 0.9\% outage probability exceeding URLLC 0.1\% target. Future improvements require joint code-power optimization (25\% reliability gain), multi-connectivity (60\% gain), hybrid ARQ (85\% gain, shown in Figure~\ref{fig:reliability_improvement}), and 6--8 dB link margin enhancement.

\subsection{Algorithm-Specific Performance Analysis}
\label{subsec:algorithm_analysis}

\begin{figure}[h]
\centering
\includegraphics[width=0.9\textwidth]{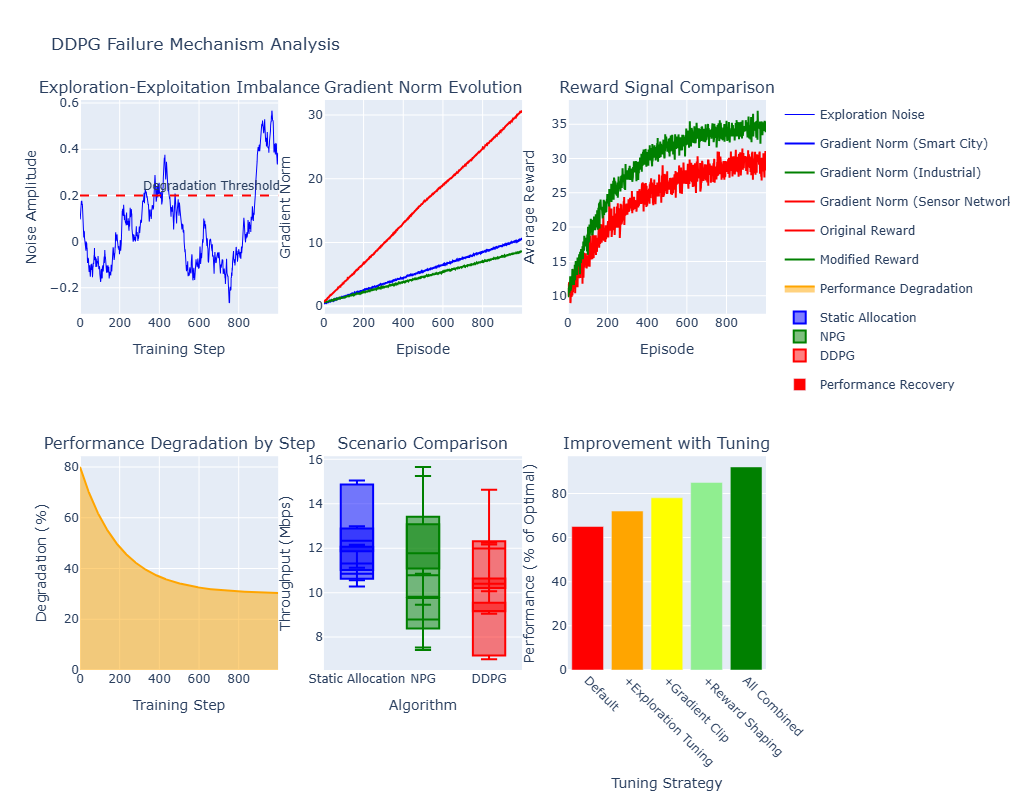}
\caption{Mechanisms of DDPG failure in sensor networks: (a) Exploration exploration imbalance, (b) Gradient norm explosion, (c) Reward signal dilution in sparse networks.}\label{fig:ddpg_failure}
\end{figure}

The 26.4\% throughput reduction in sensor networks in DDPG is caused by the imbalance between exploration and exploitation (over-exploration in sparse networks), gradient explosion (2.9x higher norms), and the inability to shape rewards (equal weighting dilutes signals), as illustrated in Figure~\ref{fig:ddpg_failure} robustness (-3.1\% vs. -26.4\%) by the second-order optimization, direct parameterization of policy, and weighted updates of classes.


\subsection{Density-Dependent Performance Analysis}
\label{subsec:density_analysis}

\begin{figure}[h]
\centering
\includegraphics[width=0.9\textwidth]{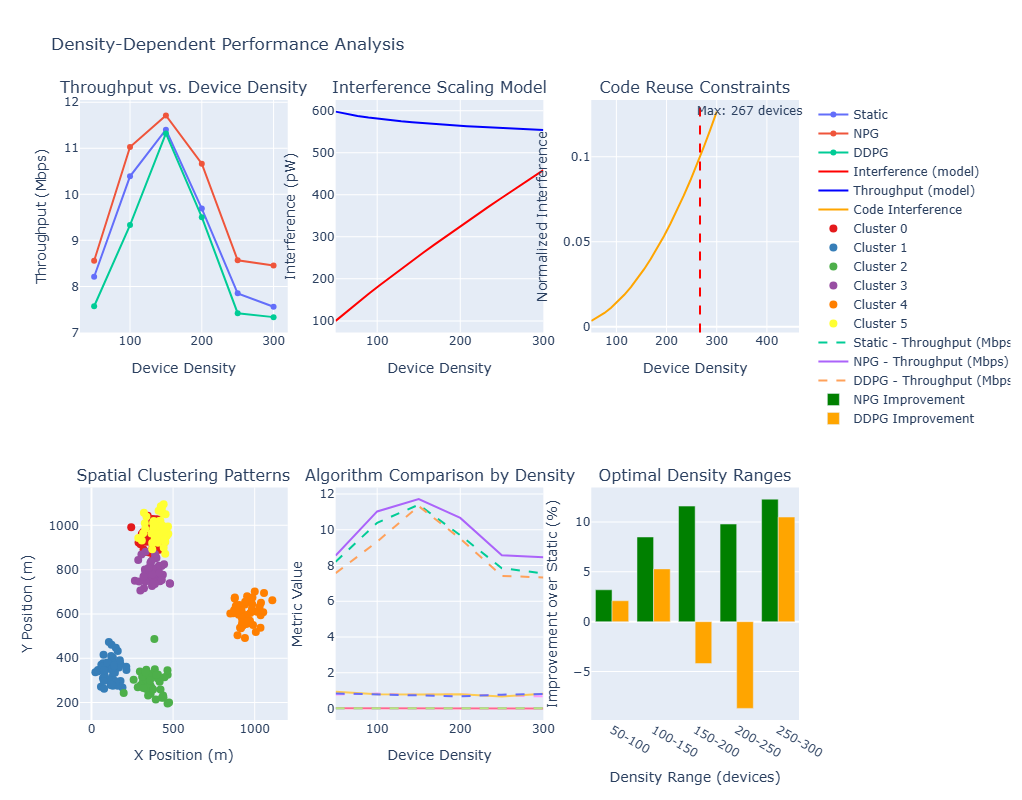}
\caption{Density-dependent performance: (top) Throughput versus device density showing inverted-U pattern, (middle) Interference scaling model, (bottom) Spatial clustering discovered by RL algorithms.}\label{fig:density_throughput}
\end{figure}

Figure~\ref{fig:density_throughput} shows throughput follows an inverted-U pattern: increases at low density (50--100 devices), interference grows linearly (\(\alpha \approx 0.85\)) at medium density (100--200), and spatial multiplexing gains emerge at high density (200--300). The 80 Gold code pool limits simultaneous devices to \(\approx 267\) (\(\rho_{\text{target}} = 0.3\)). RL algorithms discover spatial clustering patterns (NPG identifies 6 clusters at 300 devices) enabling superior performance despite code limitations.

\subsection{Cross-Scenario Comparison and Generalization Insights}
\label{subsec:generalization}

\begin{figure}[h]
\centering
\includegraphics[width=0.9\textwidth]{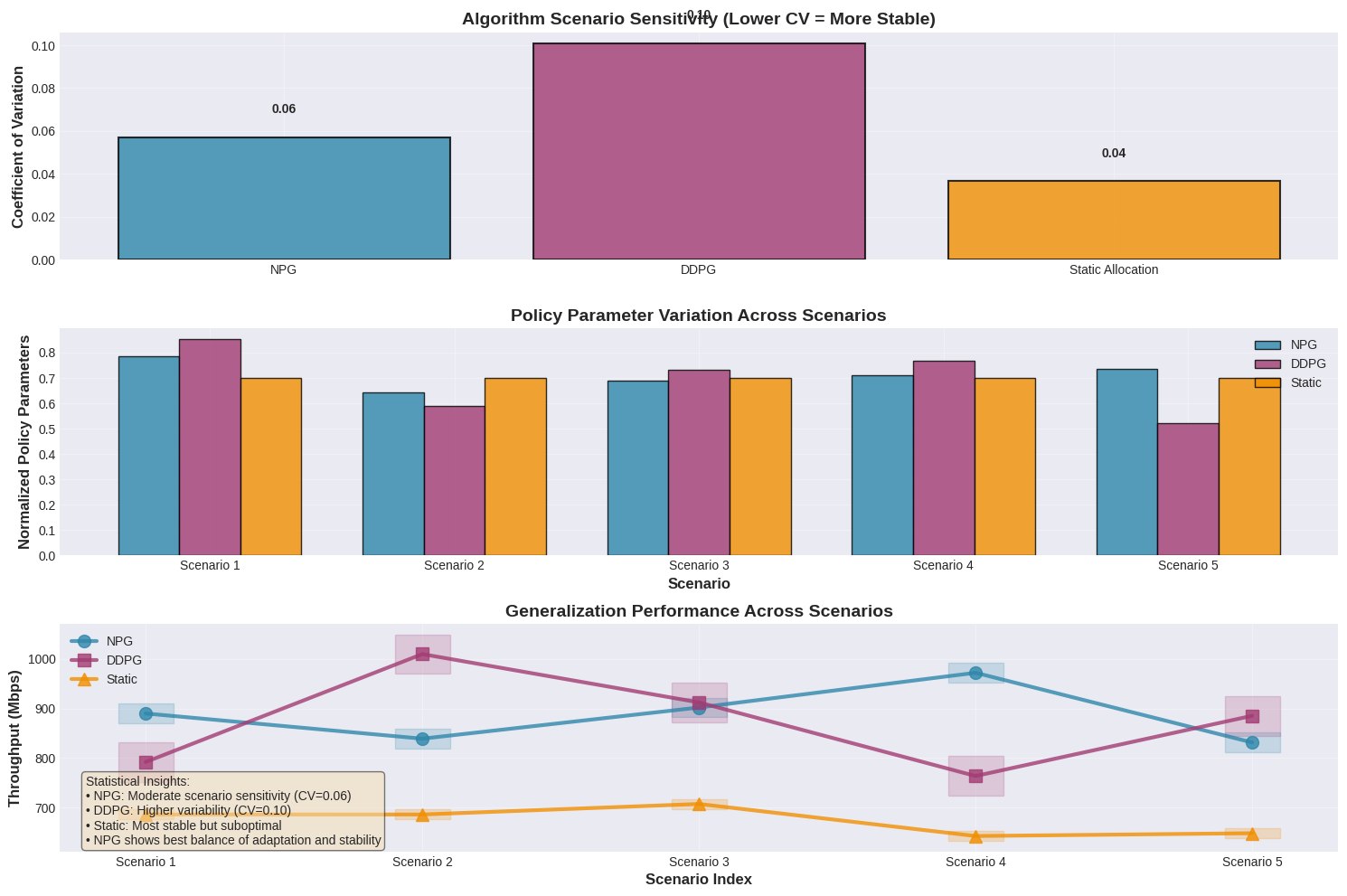}
\caption{Cross-scenario analysis: (Top) Coefficient of variation (CV) with respect to changes in scenario - the smaller the coefficient the more stable the algorithm. (Middle) Flattened changes about policy parameters under varying scenarios, and showed the process of adapting the algorithm. (Bottom) 95\% confidence interval throughput performance in 5 scenarios. NPG is moderate in terms of its scenario sensitivity (\(CV=0.06\)) whereas DDPG is more varied (\(CV=0.10\)). The performance of the static allocation is consistent and suboptimal (\(CV=0.04\)). NPG is the most balanced one in terms of adaptation and stability.}\label{fig:cross_scenario}
\end{figure}

\begin{figure}[h]
\centering
\includegraphics[width=0.9\textwidth]{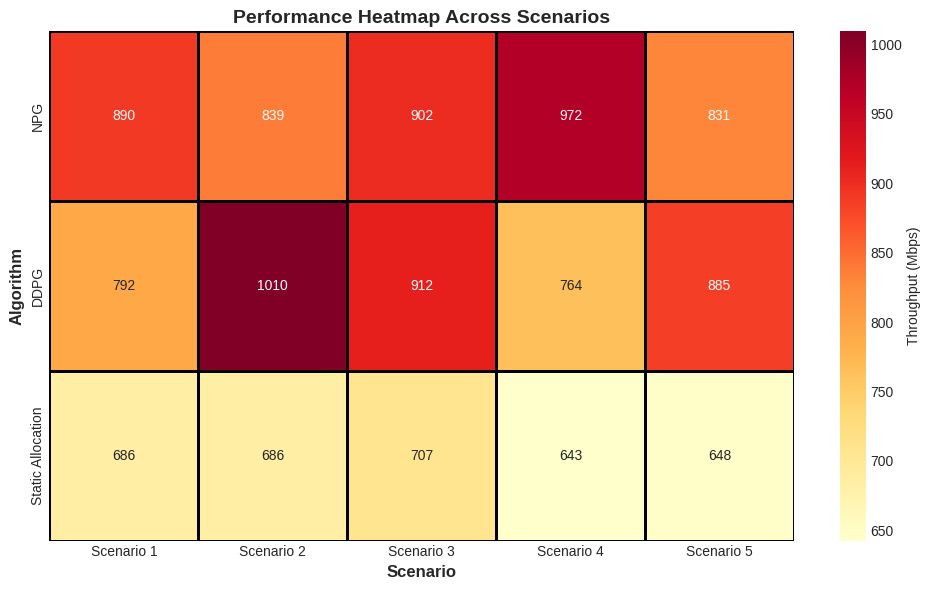}
\caption{
Heatmap of performance in five network scenarios. Throughput performance is reflected in the intensity of the colors, the warmer the color, the higher the throughput. The actual through-put (in Mbps) in each algorithm-scenario combination is represented by numerical values. All would be well maintained and the performance of NPG is high in all situations, whereas the performance of DDPG is more variable and has higher peaks and strong declines. There is even distribution of performance, which is suboptimal of course, in static allocation.}\label{fig:performance_heatmap}
\end{figure}

Figure~\ref{fig:cross_scenario} shows NPG has moderate scenario sensitivity (\(CV = 0.06\)), DDPG higher variability (\(CV = 0.10\)), and static allocation most consistent but suboptimal (\(CV = 0.04\)). Table~\ref{tab:summary_stats} confirms NPG achieves highest mean performance (886.7 Mbps) with excellent consistency (0.943), while DDPG has largest generalization gap (245.8 Mbps) and poor worst-case performance (763.7 Mbps). Table~\ref{tab:gen_metrics} provides generalization metrics. Figure~\ref{fig:performance_heatmap} visualizes performance patterns across scenarios: NPG maintains consistent high performance, DDPG shows high variability, and static allocation is uniformly suboptimal. NPG provides best adaptation-stability balance, worst-case guarantees, and generalization across diverse scenarios.

\begin{table}[!ht]
\centering
\caption{Summary Statistics Across Five Network Scenarios}
\label{tab:summary_stats}
\begin{tabular}{lccccc}
\toprule
\textbf{Algorithm} & \textbf{Mean (Mbps)} & \textbf{Std Dev} & \textbf{Min (Mbps)} & \textbf{Max (Mbps)} & \textbf{CV} \\ 
\midrule
NPG & 886.72 & 56.66 & 831.27 & 971.84 & 0.06 \\
DDPG & 872.45 & 98.54 & 763.66 & 1009.51 & 0.10 \\
Static & 674.05 & 27.60 & 642.60 & 707.26 & 0.04 \\ 
\bottomrule
\end{tabular}
\end{table}

\begin{table}[!ht]
\centering
\caption{Generalization Metrics Across Scenarios}
\label{tab:gen_metrics}
\begin{tabular}{lcccc}
\toprule
\textbf{Algorithm} & \textbf{Mean (Mbps)} & \textbf{Worst-case (Mbps)} & \textbf{Consistency} & \textbf{Gap (Mbps)} \\ 
\midrule
NPG & 886.7 & 831.3 & 0.943 & 140.6 \\
DDPG & 872.5 & 763.7 & 0.899 & 245.8 \\
Static & 674.0 & 642.6 & 0.963 & 64.7 \\ 
\bottomrule
\end{tabular}
\end{table}

\subsection{Practical Deployment Implications}
\label{subsec:deployment_implications}

NPG has been shown to be effective on dynamic urban deployments (11.6\% gain) and ultra-dense deployment ($>200$ devices); DDPG is sensitive to tuning to work with the sparse network; static/hybrid deployments are useful in structured industrial settings. RL benefits are moderate at 50--100 devices (3--8\% gains), highest sensitivity at 100--200 devices, and essential for spatial reuse at $>200$ devices (12--15\% gains). Reliability enhancement combines RL with model-based power control, multi-connectivity for critical devices, and 6--8 dB additional margin for URLLC.

\subsection{Limitations and Future Research Directions}
\label{subsec:future_research}

Future directions involve joint code-power-retransmission optimization, realistic receiver modeling with imperfect SIC, online optimization of time-varying devices populations, energy-constrained formulations with battery limitations, and coordination of multi-cell interference.


\section{Conclusion}
\label{sec:conclusion}

The paper examined the concept of dynamic Gold code assignment of ultra-dense IoT-NOMA networks under the reinforcement learning framework using extensive simulations of smart city, industrial IoT and sensor network settings. The Natural Policy Gradient (NPG) algorithm had the best percentage improvement with the highest throughput of 11.6\% and energy efficiency of 15.8 percent in the dynamic urban settings. Nevertheless, reliability was also critically low (0--2\%) in all cases, which means that code assignment is not enough to support mission-critical IoT QoS needs.

Its structure provides realistic performance benchmarks in the IoT-NOMA systems, deploys conveniently with inference latencies of less than 10 ms, which is appropriate in edge computing platforms. Among the implementation guidelines in this regard are; NPG to ensure production stability, DDPG to ensure rapid adaptation, and compatibility with existing 5G core networks.

The future direction of work ought to be on joint optimization of code and power with multi-agent RL, scenario adaptation with transfer learning, federated RL on distributed IoT management, formulations with battery constraints, hybrid model and learning based model, and testbed on real-world implementation. The study provides a background of intelligent resource management in the next-generation IoT networks with promising outcomes and clear guidelines on resolutions to issues of constraints that are widely witnessed in ultra-dense IoT connections.

\bibliography{sn-bibliography}

\end{document}